\documentclass[aps,prb,twocolumn,amsmath,amssymb,superscriptaddress,reprint,longbibliography]{revtex4-2}

\usepackage[utf8]{inputenc}
\usepackage{graphicx}
\usepackage[colorlinks=true,citecolor=blue]{hyperref}
\usepackage{units}
\newcommand{\iu}{{i\mkern1mu}}
\newcommand{\bra}[1]{\langle #1|}
\newcommand{\ket}[1]{|#1\rangle}

\renewcommand{\vec}[1]{\mathbf{#1}}

\newcommand{\kBT}{k_\text{B}T}

\newcommand{\GS}{\Gamma_\text{S}}
\newcommand{\GN}{\Gamma_\text{N}}
\newcommand{\eA}{\epsilon_\text{A}}

\begin{document}
\title{Pair-amplitude dynamics in strongly-coupled superconductor-quantum dot hybrids}
\author{Markus Heckschen}
\affiliation{Theoretische Physik, Universität Duisburg-Essen and CENIDE, D-47048 Duisburg, Germany}
\author{Björn Sothmann}
\affiliation{Theoretische Physik, Universität Duisburg-Essen and CENIDE, D-47048 Duisburg, Germany}
\date{\today}

\begin{abstract}
We consider a three-terminal system consisting of a quantum dot strongly coupled to two superconducting reservoirs in the infinite gap limit and weakly coupled to a normal metal. Using a real-time diagrammatic approach, we calculate the dynamics of the proximity-induced pair amplitude on the quantum dot. We find that after a quench the pair amplitude shows pronounced oscillations with a frequency determined by the coupling to the superconductors. In addition, it decays exponentially on a time scale set by the coupling to the normal metal. Strong oscillations of the pair amplitude occur also when the system is periodically driven both in the adiabatic and fast-driving limit. We relate the dynamics of the pair amplitude to the Josephson and Andreev current through the dot to demonstrate that it is an experimentally accessible quantity.
\end{abstract}

\maketitle
\section{Introduction}
Superconductivity is an active field of research in modern condensed-matter physics. It is of interest from a fundamental perspective because it provides an example of quantum coherence at the macroscopic scale which gives rise to phenomena such as the Josephson effect where a dissipationless charge current flows without an applied bias voltage~\cite{josephson_possible_1962}. Furthermore, it provides an important ingredient for applications such as superconducting qubits in future quantum computers~\cite{clarke_superconducting_2008,arute_quantum_2019,kjaergaard_superconducting_2020}.

The transition from a normal metal into the superconducting state is a second-order phase transition in which the U(1) symmetry is spontaneously broken by the condensation of electrons into spin-singlet, $s$-wave Cooper pairs as has been established by the microscopic BCS theory of superconductivity by Bardeen, Cooper and Schrieffer~\cite{bardeen_theory_1957}. The associated order parameter is a complex number $\Delta e^{i\phi}$ which corresponds to the macroscopic wave function of Cooper pairs. 
The order parameter is a dynamic quantity which can exhibit collective oscillations such as the Nambu-Goldstone mode and the Higgs mode. The former is a gapless excitation of the superconducting phase $\varphi$ while the latter is a gapped excitation of the absolute value $\Delta$ of the superconducting order parameter with minimal excitation energy $2\Delta$. Due to the Anderson-Higgs mechanism, the frequency of the Nambu-Goldstone mode is shifted to the plasma frequency in bulk superconductors~\cite{anderson_coherent_1958,nambu_quasi-particles_1960,anderson_plasmons_1963}. Therefore, the Higgs mode is the only collective low-energy excitation of the superconducting order parameter and stable against the decay into other modes.

The experimental detection of the Higgs mode is challenging for various reasons. The Higgs mode is charge neutral and couples only quadratically to external electromagnetic fields. Therefore, large field strengths are needed to excite it. The minimal excitation energy of the Higgs mode in typical BCS superconductors is in the THz range where suitable source and detectors have been developed only recently. Furthermore, one has to exclude the additional excitation of quasiparticles by Cooper pair breaking which requires the same energy as the excitation of the Higgs mode.
The first detection of the Higgs mode was reported in Refs.~\cite{sooryakumar_raman_1980,sooryakumar_raman_1981} in materials where superconductivity coexists with a charge-density wave. The latter couples the Higgs mode to phonons such that it can be detected as an additional peak in the Raman spectrum. The advent of powerful Thz lasers made it possible to directly excite and observe the Higgs mode by pump-probe spectroscopy~\cite{matsunaga_higgs_2013}. The experimentally observed oscillation of the electromagnetic response can be explained theoretically in terms of the Anderson pseudospin dynamics in a two-dimensional BCS model~\cite{chou_twisting_2017} and by a gauge-invariant kinetic equation of superconductivity~\cite{yang_gauge-invariant_2019}. The Higgs mode has also been probed by its third-order nonlinear optical response which provides a clear distinction between the Higgs mode and quasiparticle excitations~\cite{matsunaga_light-induced_2014, tsuji_theory_2015}. Recently, the Higgs mode has also been detected in a strongly-interacting fermionic superfluid~\cite{behrle_higgs_2018}. Further theoretical studies focused on the Higgs mode in unconventional $d$-wave superconductors~\cite{schwarz_classification_2020, chu_phase-resolved_2020} and on the coupling of the Higgs and Leggett modes in two-gap superconductors~\cite{krull_coupling_2016}. Current reviews on the Higgs mode in superconductors can be found in Refs.~\cite{pekker_amplitude/higgs_2015, shimano_higgs_2020}.

So far, the Higgs mode has mainly been studied in bulk superconductors. Recently, it was demonstrated that the leakage of quasiparticles into a normal conductor leads to a strong damping of the Higgs mode in a superconductor-normal metal junction. In addition, the coherent tunneling of electrons across the junction gives rise to the occurrence of two new Higgs modes~\cite{vadimov_higgs_2019}. Signatures of the Higgs mode in the charge current of a superconductor-normal metal tunnel junction were reported in Ref.~\cite{tang_signatures_2020}.
Superconductors can induce superconducting correlations in nearby nonsuperconducting materials via the proximity effect. This raises the interesting question about the nature of the dynamics of the proximity-induced pair amplitude in the nonsuperconducting part of the junction. So far, this dynamics has been studied in a weakly coupled, temperature-biased  superconductor-quantum dot hybrid~\cite{kamp_higgs-like_2021}. It was shown that both the absolute value and the phase of the induced pair amplitude can oscillate with time. However, a strong damping due to tunneling of quasiparticles typically overcomes the oscillations unless the system is driven externally. 
This motivates us to theoretically investigate here the pair amplitude dynamics in a quantum dot strongly coupled to superconductors in the regime where the superconducting gap is much larger than temperature. In this parameter regime, tunneling of quasiparticles between the dot and the superconductors is exponentially suppressed while at the same time a strong proximity effect can occur. We account for dissipative processes by weakly coupling the quantum dot to an additional normal metal reservoir. 
The charge dynamics of this system for pulsed driving has been previously studied in light of  potential applications in quantum computing~\cite{moghaddam_driven_2012}.
Using a real-time diagrammatic approach which enables us to account for the proximity effect and strong Coulomb interactions on the dot in a nonequilibrium scenario, we study the resulting pair amplitude dyamics both after a quench as well as under a periodic driving of the system. We find that in both scenarios pronounced oscillations of the pair amplitude occur, thus overcoming the limitations of the weak-coupling limit~\cite{kamp_higgs-like_2021}.

The paper is organized as follows. The theoretical model of our hybrid system is presented in Sec.~\ref{sec:model}. In Sec.~\ref{sec:rtd}, we introduce our theoretical description based on a real-time diagrammatic approach. We discuss our results for the dynamics after a quench and under periodic driving in Secs.~\ref{sec:quench} and~\ref{sec:period}, respectively. Conclusions are drawn in Sec.~\ref{sec:conclusion}.

\section{\label{sec:model}Model}
\begin{figure}
	\includegraphics[width=\columnwidth]{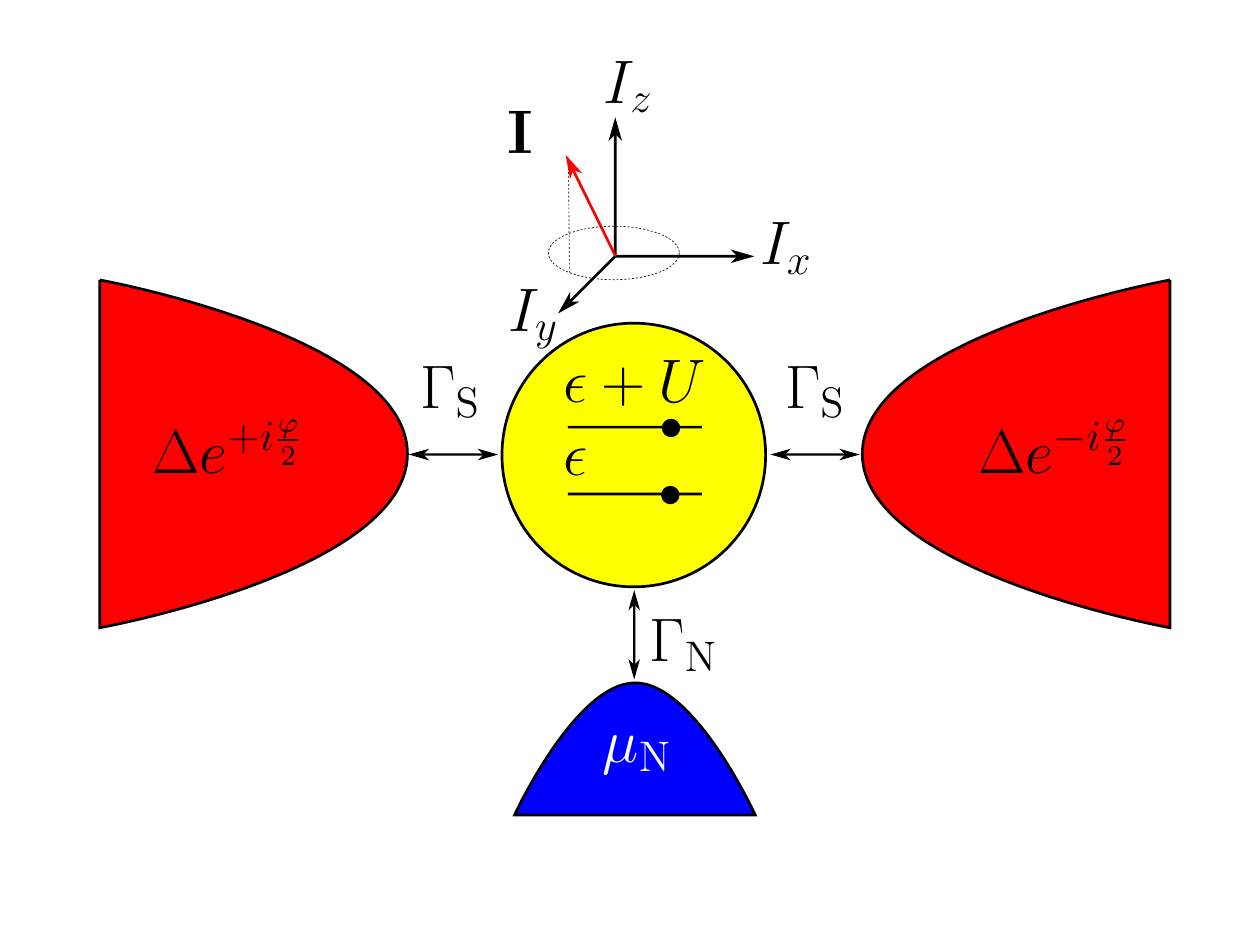}
	\caption{Schematic sketch of the three-terminal system. A single-level quantum dot is strongly coupled to two grounded superconductors with superconducting phases $\varphi_\eta=\pm\varphi/2$ by tunnel coupling strengths $\Gamma_\text{SL}=\Gamma_\text{SR}=\GS$. Furthermore, the quantum dot is weakly tunnel coupled to a normal conductor with electrochemical potential $\mu_\text{N}$. The proximity effect induces superconducting correlations on the quantum dot which are characterized by the pseudospin $\vec I$.} 
	\label{fig::sys}
\end{figure}
We consider a three-terminal setup composed of two conventional BCS superconductors coupled symmetrically to a single-level quantum dot which in addition is weakly coupled to a normal conductor; see Fig.~\ref{fig::sys}. The coupling to the superconductors induces superconducting correlations on the quantum dot via the proximity effect while the coupling to the normal metal allows for the tunneling of single electrons which in turn gives rise to dissipation.

The setup is described by the total Hamiltonian
\begin{equation}
	H = \sum_\eta H_\eta + H_\text{dot} + H_\text{tun}.
\end{equation}
The first term describes the three leads $\eta\in\{\text{SL},\text{SR},\text{N}\}$ in terms of a mean-field BCS-Hamiltonian
\begin{equation}
	H_\eta = \sum_{\mathbf{k}\sigma} \epsilon_{\eta,\mathbf{k}} a_{\eta \mathbf{k}\sigma}^\dagger a_{\eta \mathbf{k}\sigma} + \sum_\mathbf{k}\left(\Delta_\eta e^{\iu \varphi_\eta} a_{\eta \mathbf{k}\uparrow}a_{\eta-\mathbf{k}\downarrow} + \text{H.c.}\right).
\end{equation} 
The first term corresponds to the kinetic energy of electrons in lead $\eta$ with spin $\sigma$ and momentum $\mathbf{k}$. 
The second term describes the superconducting pairing and vanishes for the normal lead.
Here, $\Delta$ is the absolute value of the superconducting order parameter which we assume to be equal in both superconductors and which is zero for the normal lead. 
The phases of the superconducting order parameters are chosen symmetrically as $\varphi_\text{L}=-\varphi_\text{R}=\varphi/2$ without loss of generality.
Both superconductors are assumed to have the same critical temperature and are grounded, $\mu_\text{S}=0$. 
The system can be driven into a nonequilibrium state by applying a bias voltage $\mu_\text{N}=e V$ to the normal reservoir.

The single-level, spin-degenerate quantum dot with level position $\epsilon$ is described by the dot Hamiltonian
\begin{equation}
	H_\text{dot} = \sum_\sigma \epsilon c_\sigma^\dagger c_\sigma + Uc_\uparrow^\dagger c_\uparrow c_\downarrow^\dagger c_\downarrow,
\end{equation}
where the second terms denotes the on-site Coulomb repulsion $U$.
The level position can be tuned by a gate voltage.

The tunneling Hamiltonian is given by
\begin{equation}
	H_\text{tun} = \sum_{\eta \mathbf{k}\sigma} t_\eta a_{\eta \mathbf{k}\sigma}^\dagger c_\sigma + \text{H.c.},
\end{equation}
with tunnel-matrix element $t_\eta$ which we assume to be spin and momentum independent. 
The matrix elements are related to the tunnel-coupling strength $\Gamma_\eta=2\pi |t_\eta|^2\rho_\eta^\text{N}$, with density of states in the normal state $\rho_\eta^\text{N}$.
Due to the symmetric coupling to the superconductors,  we have $\Gamma_{\text{SL}}=\Gamma_{\text{SR}}=\GS$.

In this work, we concentrate on the dynamics of the proximity-induced superconducting pair amplitude on the quantum dot. As it describes the dynamics of Cooper pairs inside the superconducting gap, our focus is on the subgap physics on the system at hands. Therefore, it is a reasonable approximation to consider our setup in the infinite-gap limit where the superconducting gap is assumed to be the largest energy scale in the problem. In this limit, we can take into account the coupling between the dot and the superconducting reservoirs exactly since even- and odd-parity dot states decouple and the system can be mapped onto a noninteracting one with renormalized level positions~\cite{rozhkov_interacting-impurity_2000}.

The exact resummation results in an effective dot Hamiltonian $H_\text{eff}=H_\text{dot}-H_\text{p}$ where for a symmetric coupling to the superconductors the pairing Hamiltonian is given by
\begin{equation}
	H_\text{p}=\GS \cos\left(\frac{\varphi}{2}\right)(c_\downarrow c_\uparrow+  c_\uparrow^\dagger c_\downarrow^\dagger).
\end{equation} 
The eigenbasis of $H_\text{eff}$ is given by $\ket{\uparrow}$, $\ket{\downarrow}$ which describe the singly-occupied the dot with an spin-up or spin-down electron and 
\begin{equation}
	\ket{\pm}= \frac1{\sqrt{2}}\left[\mp \sqrt{1\mp\frac{\delta}{2\eA}}\ket{0} + \sqrt{1\pm\frac{\delta}{2\eA}}\ket{d}\right],
\end{equation}
which describe a coherent superposition of empty and double-occupied dot states, $\ket{0}, \ket{d}$ with  $\eA= \sqrt{\left(\frac{\delta}{2}\right)^2+|\chi|^2}$, the level detuning $\delta=2\epsilon+U$ which characterizes deviations from the particle-hole symmetric point $\delta=0$, and eigenenergies $E_\sigma=\epsilon$ and $E_\pm=\frac{\delta}{2}\pm\epsilon_\text{A}$~\cite{futterer_renormalization_2013, rozhkov_interacting-impurity_2000}. 
The excitation energies of the quantum dot are given by the differences of the eigenenergies $\pm(E_\pm-E_\sigma)$ and can be identified as the Andreev bound state energies in the infinite-gap limit
\begin{equation}
	E_{\text{A},\gamma',\gamma}=\gamma'\frac{U}{2}+\gamma\epsilon_\text{A}\quad \gamma',\gamma\in\{\pm1\}. \label{eq::abs}.
\end{equation}
The superposition of the zero-occupied $\ket{0}$ and double-occupied dot $\ket{d}$ becomes largest for $\delta=0$. For large detunings, the states $\ket{\pm}$ are approximately given by the empty and doubly-occupied state.

\section{\label{sec:rtd}Real-time diagrammatics}
In order to describe the pair-amplitude dynamics on the quantum dot, we make use of a real-time diagrammatic approach~\cite{konig_zero-bias_1996,konig_resonant_1996} in an extension to superconducting reservoirs~\cite{governale_real-time_2008,*governale_erratum:_2008}. The key idea of the real-time diagrammatics is to integrate out all noninteracting degrees of freedom and to describe the quantum dot with a reduced density matrix $\rho_\text{red.}$.
Our method allows us to include arbitrary interactions on the quantum dot. Furthermore, it captures the coupling between the dot and the superconductors exactly in the infinite-gap limit. The coupling to the normal lead is accounted for in a systematic perturbation theory in the tunnel coupling $\Gamma_\text{N}$. In the following, we restrict ourselves to processes up to first order in $\Gamma_\text{N}$ as they dominate transport in the parameter regimes considered later on.

The matrix elements of the reduced density matrix are given by
\begin{equation}
	P_{\chi_2}^{\chi_1}=\bra{\chi_1}\rho_\text{red}\ket{\chi_2}.
\end{equation}
In order to properly describe the dynamics of the pair amplitude, we have to take into account the diagonal density matrix elements $P^\chi_\chi$ which describe the probability to find the dot in the state $\chi$ as well as the offdiagonal elements $P_+^-$ and $P_-^+$. The latter are crucial to capture the fast coherent tunneling of Cooper pairs which occurs on a time scale given by $\Gamma_\text{S}$~\cite{rajabi_waiting_2013}.

The time evolution of the reduced density matrix elements is given by a generalized master equation of the form
\begin{equation}
	\frac{d}{dt}P_{\chi_2}^{\chi_1}(t)=-\frac{\iu}{\hbar}(E_{\chi_1}-E_{\chi_2})+\sum_{\chi_1'\chi_2'}\int_{-\infty}^t dt' W_{\chi_2\chi_2'}^{\chi_1\chi_1'}(t,t')P_{\chi_2'}^{\chi_1'}(t').
\end{equation}
The first term on the right side of the master equation describes the coherent evolution of the quantum dot. In particular, it accounts for the coherent Cooper pair dynamics due to the coupling to the superconducting leads. The second term describes the dissipative coupling to the normal lead. The generalized transition rates $W_{\chi_2\chi_2'}^{\chi_1\chi_1'}$ are given by irreducible selfenergies of the quantum dot propagator on the Keldysh contour.

The generalized master equation can be transformed into a physically intuitive form by introducing a pseudospin degree of freedom which characterizes the coherent superposition of the empty and doubly occupied dot state via 
\begin{equation}
	I_x =\frac{P_0^d + P_d^0}{2}, \quad I_y = \iu\frac{P_0^d - P_d^0}{2}, \quad I_z= \frac{P_d-P_0}{2}.
\end{equation}
In the infinite-gap limit, the dynamics of the pseudospin decouples from the dot occupations and its time evolution is governed by a Bloch-type equation
\begin{equation}
	\hbar\frac{d\mathbf{I}}{dt}=\mathbf{A}-\mathbf{R}\cdot\mathbf{I}-\mathbf{B}\times\mathbf{I}.\label{eq::bloch}
\end{equation}
In Eq.~\eqref{eq::bloch}, $\mathbf{A}$ denotes the accumulation vector which describes accumulation of pseudospin on the quantum dot due to tunneling electrons. 
It is given by
\begin{equation}
	\mathbf{A}=-\frac{\GN}{4}
	\left(\begin{array}{c}
		\frac{\GS}{\eA}\cos\frac{\varphi}{2}\sum_{\gamma\gamma'=\pm}\gamma f(E_{\text{A},\gamma',\gamma}) \\
		\frac{\GS}{\pi\eA}\cos\frac{\varphi}{2}\sum_{\gamma\gamma'=\pm}\gamma\varphi(E_{\text{A},\gamma',\gamma}) \\
		\sum_{\gamma\gamma'=\pm}\left(1+\gamma\frac{\delta}{2\eA}\right)\left[\frac1{2}-f(E_{\text{A},\gamma',\gamma})\right]
	\end{array}\right),\nonumber
\end{equation}
where $f(x)=\{1+\exp [(x-\mu_\text{N})/k_\text{B}T]\}^{-1}$ denotes the Fermi function and $\varphi(x)=\text{Re}\left[\psi\left(\frac1{2}+i \frac{\omega}{2\pi \kBT}\right)\right]$ with the Digamma function $\psi(x)$.

The second term in Eq.~\eqref{eq::bloch} accounts for the anisotropic relaxation of the pseudospin in terms of the relaxation tensor $\mathbf{R}$. 
The relaxation tensor describes the decoherence in the system which arises from tunneling of single electrons between quantum dot and normal conductor.
The nonzero elements of the relaxation tensor are
\begin{align}
	R_{xx}&=R_{yy}=\frac{\GN}{2}\sum_{\gamma\gamma'=\pm}\left(1-\gamma\frac{\delta}{2\eA}\right)\left(\frac1{2}-\gamma'f(E_{\text{A},\gamma'\gamma})\right) ,\nonumber\\
	R_{zz}&=\frac{\GN}{2}\sum_{\gamma\gamma'=\pm}\left(1+\gamma\frac{\delta}{2\eA}\right)\left(\frac1{2}-\gamma'f(E_{\text{A},\gamma',\gamma})\right),\nonumber\\
	R_{xz}&=R_{zx}=\frac{\GN\GS}{2\eA}\cos\frac{\varphi}{2}\sum_{\gamma\gamma'=\pm}\gamma\gamma'f(E_{\text{A},\gamma',\gamma})\nonumber.
\end{align}

Finally, the last term in Eq.~\eqref{eq::bloch} describes the precession of the pseudospin in an effective exchange field $\mathbf{B}=\vec B_0+\vec B_1$. The exchange field has two contributions. The first one,
\begin{equation}
    \mathbf{B}_0 =\left(\begin{array}{c} 2\GS\cos\frac{\varphi}{2} \\0\\-\delta\end{array}\right),
\end{equation}
is due to the coupling to the superconductors and the level detuning of the quantum dot. The second contribution,
\begin{equation}
    \mathbf{B}_1 =-\frac{\GN}{2\pi}\left(\begin{array}{c}\frac{\GS}{\eA}\cos\frac{\varphi}{2}\sum_{\gamma,\gamma'=\pm}\gamma\gamma'\varphi(E_{\text{A},\gamma',\gamma})\\0\\\sum_{\gamma,\gamma'}\left(1-\gamma\frac{\delta}{2\eA}\right)\gamma'\varphi(E_{\text{A},\gamma',\gamma})\end{array}\right),
\end{equation}
is due to the renormalization of the dot level positions which arises from virtual tunneling processes between the dot and the normal lead. The level renormalization is an interaction-driven effect that vanishes in the limit of a noninteracting quantum dot. Similar effects are known from quantum-dot spin valves and weakly coupled superconductor-quantum dot hybrids~\cite{konig_interaction-driven_2003,kamp_phase-dependent_2019}. Since $\vec B_1$ is of first order in the tunnel coupling $\GN$, it is in general much smaller than $\vec B_0$.

The superconducting pair amplitude on the quantum dot,
\begin{equation}
	\mathcal{F}=\left<c_\downarrow c_\uparrow\right>,
\end{equation}
is closely linked to the pseudospin. In particular, the absolute value of the pair amplitude is given by the length of the projection of $\vec I$ into the $x-y$ plane,
\begin{equation}
	|\mathcal{F}|=\sqrt{I_x^2+I_y^2}.
\end{equation}
We remark that the absolute value of the pair amplitude on the dot is a dimensionless number that can take values in the range $0\leq |\mathcal F|\leq 1/2$.

The real-time diagrammatic technique also allows us to calculate the Josephson current which flows between the two superconductors as well as the Andreev current which flows in the normal conductor~\cite{governale_real-time_2008}. They are proportional to the pseudospin components $I_x$ and $I_y$, respectively,
\begin{align}
	J_\text{jos} &=\frac{2e}{\hbar}\GS I_x \sin\frac{\varphi}{2}\label{eq::Ijos},\\
	J_\text{and} &=-\frac{4e}{\hbar}\GS I_y \cos\frac{\varphi}{2}\label{eq::Iand}.
\end{align}
Thus, the dynamics of the dot's pair amplitude can be reconstructed straightforwardly from the two time-resolved currents.

\section{\label{sec:results}Results}
In this section we analyze the dynamics of the pair amplitude induced on the quantum dot by the superconducting proximity effect. 
First, we investigate the dynamics after quenching the system in Sect.~\ref{sec:quench}. Then, we turn to the case in which the system is periodically driven by an external force in Sect.~\ref{sec:period}.

\subsection{\label{sec:quench}Quench dynamics}
\begin{figure}
	\includegraphics[width=\columnwidth]{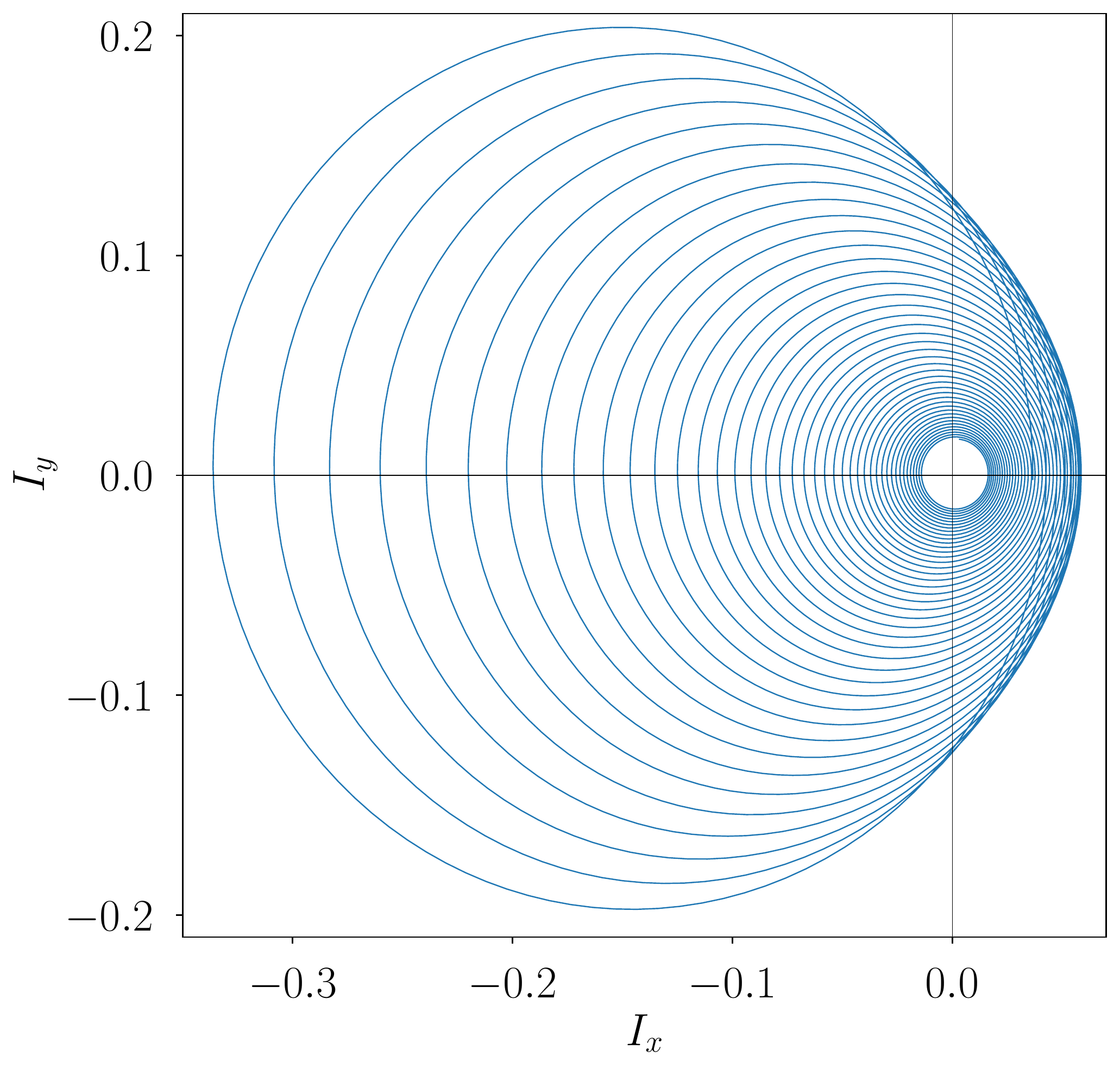}
	\caption{Pseudospin in the $x$-$y$-plane as a function of time after a quench from $\delta=-10U$ to $\delta=U/2$. The pseudospin dynamics starts close to the origin and ends after $t=\unit[2.5]{\GN/\hbar}$. The pseudospin rotates counter-clockwise around the exchange field. The other parameters for the plot are $\Gamma_S=U/10$, $\Gamma_N=5\times10^{-3}U$, $\mu_N=U/2$, $k_BT=U/100$ and $\varphi=\pi/4$.}
	\label{fig::xypseudospin}
\end{figure}
In the following, we consider the pair amplitude dynamics after a quench at time $t_0=0$ where the dot is prepared in an initial state and then relaxes into its stationary nonequilibrium state. As the accumulation vector $\vec A$, the relaxation tensor $\vec R$ and the exchange field $\vec B$ are time independent after the quench, we can formally solve the Bloch equation~\eqref{eq::bloch} by 
\begin{equation}
		\mathbf{I}(t)=\sum_{j=1}^3 e^{\lambda_j t}\mathbf{r}_j(\mathbf{l}_j)^T\cdot\left(\mathbf{I}^{(0)}-\mathbf{M}^{-1}\cdot\mathbf{A}\right)+\mathbf{M}^{-1}\cdot\mathbf{A}.\label{eq::quench}
\end{equation}
Here, $\vec I^{(0)}$ denotes the initial value of the pseudospin right before the quench. The matrix $\mathbf{M}$ describes the evolution of the pseudospin due to relaxation and rotation $M_{ij}=-R_{ij}-\sum_k\epsilon_{ijk}B_j$. It has the normalized left and right eigenvectors $\mathbf{l}_i$ and $\mathbf{r}_i$, respectively, and the corresponding eigenvalues $\lambda_j$. For sufficiently strong coupling to the superconductor, $\GS\gg\GN$, one of the eigenvalues is real while the two others are complex-conjugates of each other. The explicit form of $\mathbf{M}$ and its eigensystem are given in Appendix~\ref{app::eigensystem}.

Equation~\eqref{eq::quench} shows that the pseudospin tends to the stationary value $\vec M^{-1}\cdot \vec A$ which depends on the accumulation term, the relaxation tensor and the exchange field in a nontrivial manner. The pseudospin dynamics is driven by the difference between the initial value $\vec I^{(0)}$ and the stationary state. The dynamics shows in general two different types of time dependence. On the one hand, the pseudospin exhibits an exponential relaxation towards the stationary state on a time scale $\hbar/\GN$. In addition, there is an oscillatory time dependence which physically arises from the precession of the pseudospin around the exchange field with a frequency given by $|\vec B|$. For strong coupling to the superconductor, the precession frequency is given by $f_0=\sqrt{\delta^2+4\GS^2\cos^2\varphi/2}$ if we neglect contributions to the exchange field which are first order in $\GN$. Therefore, for a given coupling $\GS$ to the superconductors, one can manipulate the precession frequency by the superconducting phase difference $\varphi$ and the detuning $\delta$. We remark, however, that for a detuning much larger than $\GS$ the superconducting proximity effect becomes suppressed which provides an upper bound on the possible oscillation frequency.

In the following, we illustrate the quench dynamics for the situation where the dot is prepared in an initial state with large detuning $\delta=-10U$. In consequence, the dot is mostly doubly occupied and the initial pseudospin is nearly parallel to the $z$ axis, $\mathbf{I}^{(0)}\approx(0,0,1/2)$. Subsequently, the detuning is changed to $\delta=U/2$ and the resulting time evolution of the pseudospin is calculated. Figure~\ref{fig::xypseudospin} shows the projection of the pseudospin dynamics onto the $x-y$ plane. The projected dynamics starts close to the origin and then precesses around the exchange field while slowly relaxing due to tunneling processes between the dot and the normal lead. For the parameters chosen, the projected dynamics exhibits a nearly circular trajectory. We remark that for stronger couplings to the superconductor, the projection becomes elliptical due to the larger $x$ component of the exchange field.

\begin{figure}
	\includegraphics[width=\columnwidth]{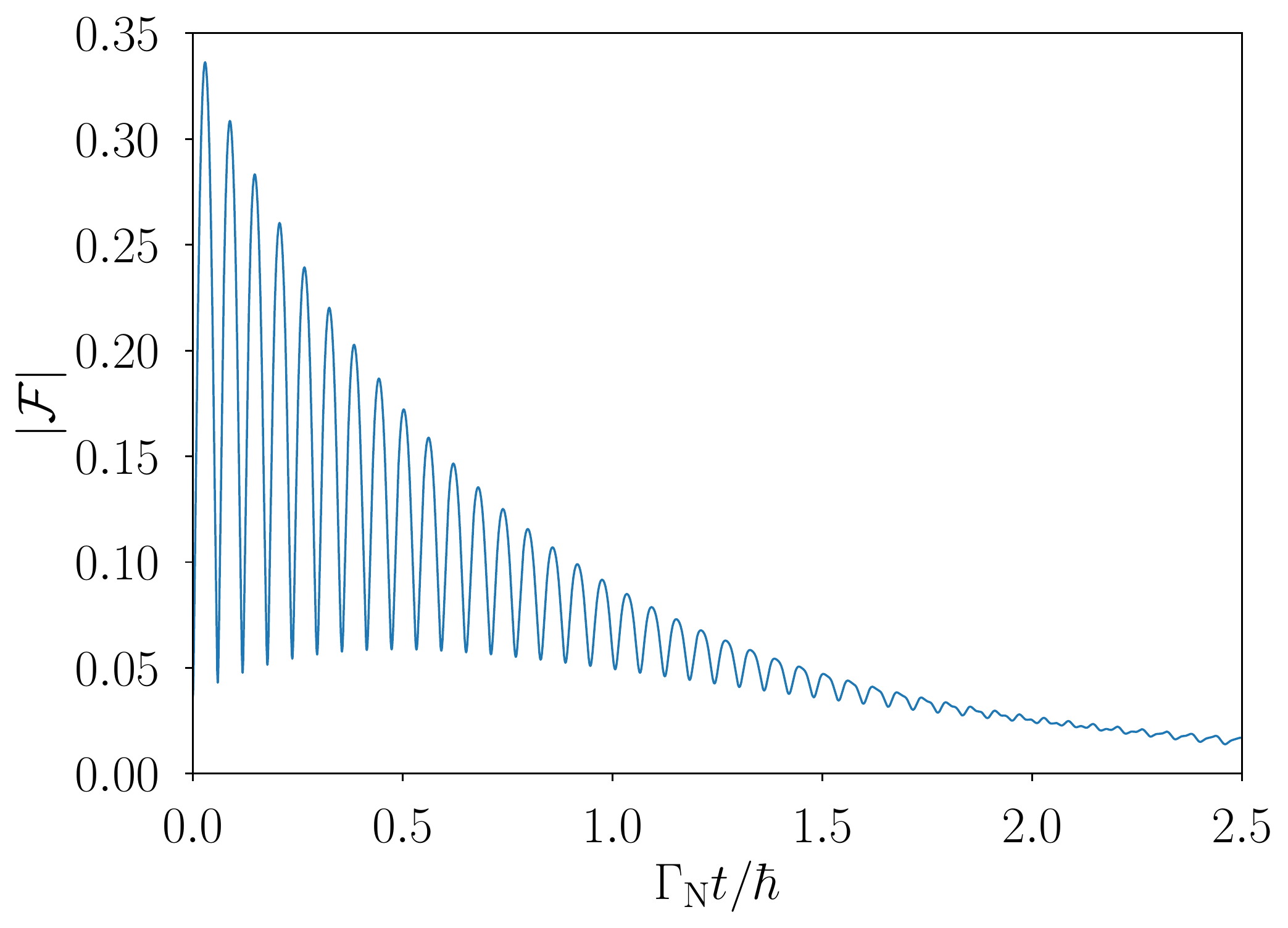}
	\caption{Pair amplitude as a function of time after a quench. Parameters are the same as in Fig.~\ref{fig::xypseudospin}.}
	\label{fig::quenchpair}
\end{figure}

We now turn to the dynamics of the absolute value of the superconducting pair amplitude $|\mathcal F|$ on the dot which is given by the length of the projection of the pseudospin onto the $x-y$ plane and which is shown in Fig.~\ref{fig::quenchpair}. Right at the quench, $|\mathcal F|$ is small but quickly rises to its maximal value on a time scale given by $\hbar/\GS$ due to the coherent tunneling of Cooper pairs between the dot and the strongly coupled superconductors. The coherent Cooper pair tunneling subsequently gives rise to pronounced oscillations of the dot's pair amplitude. In addition, both the average value of the absolute value of pair amplitude as well as its oscillation amplitude decay exponentially with time. The time scale for this decay is given by $\hbar/\GN$ and arises from the tunneling of single electrons between the normal lead and the proximitized quantum dot.

\begin{figure}
	\includegraphics[width=\columnwidth]{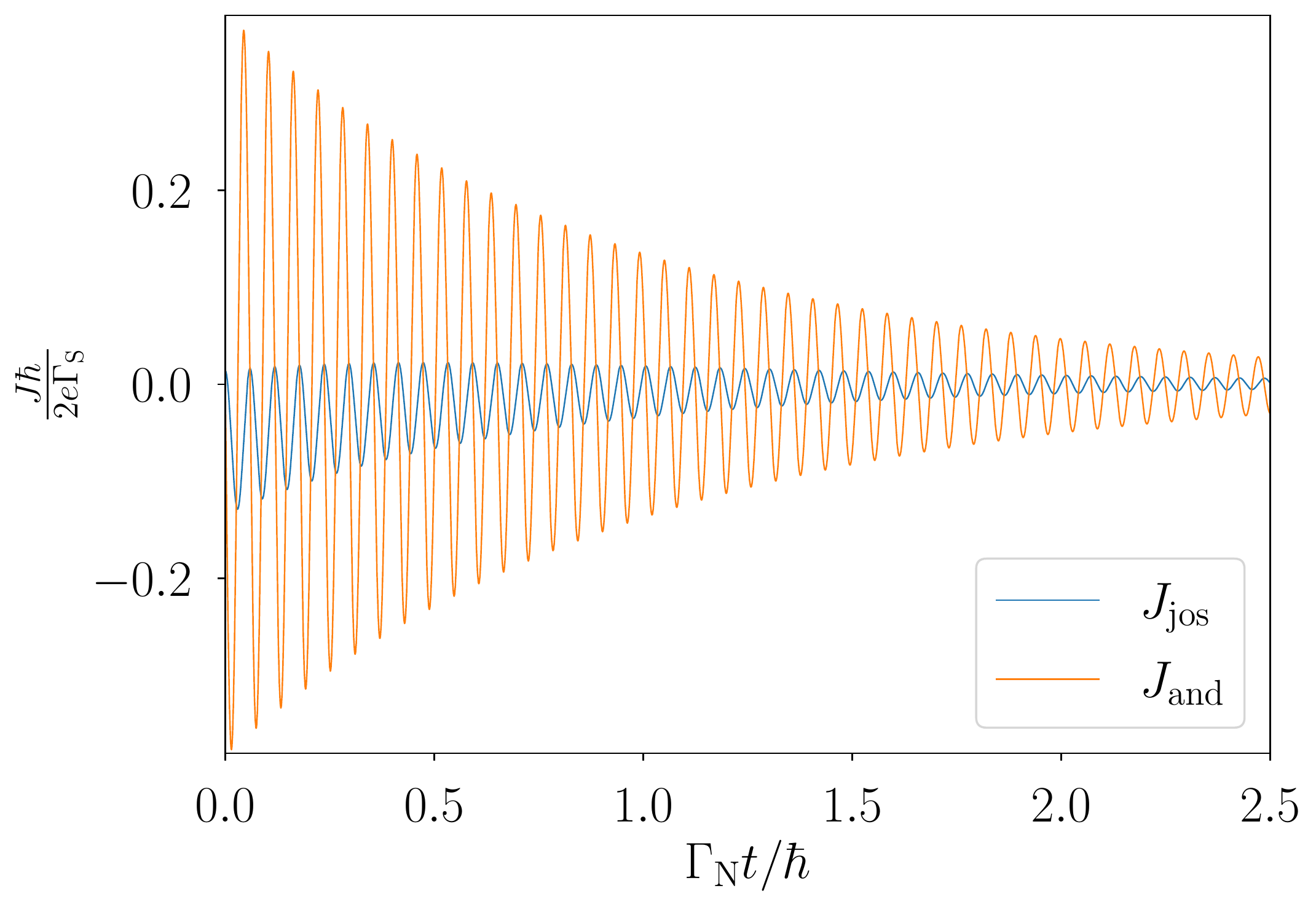}
	\caption{Josephson and Andreev current as a function of time after a quench. Parameters are the same as in Fig.~\ref{fig::xypseudospin}.}
	\label{fig::currents}
\end{figure}

Finally, we turn to the time-dependent currents in the system which, as can be seen in Eqs.~\eqref{eq::Ijos} and~\eqref{eq::Iand} directly reflect the pseudospin and, therefore, the pair-amplitude dynamics on the quantum dot. Figure~\ref{fig::currents} shows the Josephson and Andreev current as a function of time. As expected, both currents show an interplay between fast oscillations on the time scale $\hbar/\GS$ and an exponential decay on the time scale $\hbar/\GN$ just like the pseudospin and the pair amplitude. Hence, measuring both the time-resolved Josephson and Andreev current through the quantum dot provides direct access to the pair amplitude dynamics. For a coupling strength of $\GS/\hbar\sim\unit[1]{GHz}$ the currents are expected to be in the sub-\unit{nA} regime.

We conclude our analysis of the quench dynamics by remarking that in our system the frequency of the coherent pair amplitude oscillations and the time scale of its exponential decay are independent of each other as they arise from different physical processes. This allows us to explore a rich pair-amplitude dynamics which goes beyond the Higgs mode in bulk superconductors. In particular, we also find that a quantum dot strongly coupled to a large-gap superconductor is favorable to observe its pair amplitude dynamics over a quantum dot weakly coupled to a finite-gap superconductor where coherent oscillations and relaxation occur on the same time scale~\cite{kamp_higgs-like_2021}.

\subsection{\label{sec:period}Periodic driving}
In the following, we discuss to what extent the pair amplitude responds to an external parameter driving. The driving provides an additional energy intake that can help to overcome the damping of pair amplitude oscillations. This allows us to study the pair amplitude over a longer time than in the case of a quench. 
We focus on driving in the two limiting cases of adiabatic and fast driving. On the one hand, we will analyze a time-dependent driving of the phase difference between the superconductors $\varphi(t)=2\omega t$ and on the other hand a time-dependent detuning $\delta(t)=\delta_0+\delta_1 \cos(\omega t)$.  

\subsubsection{\label{sec:adiabatic}Adiabatic Driving}
\begin{figure}
	\includegraphics[width=\columnwidth]{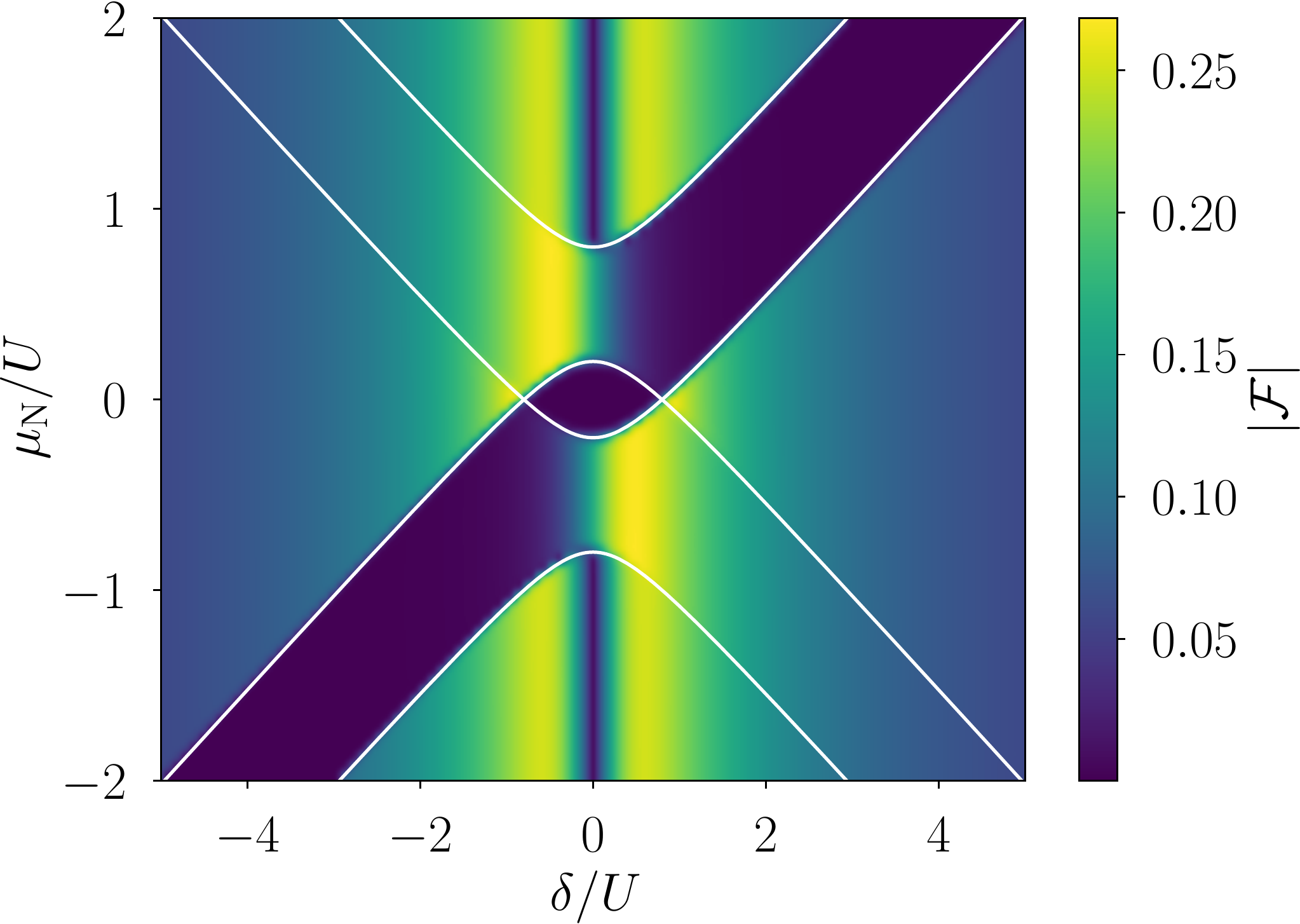}
	\caption{Pair amplitude as a function of the chemical potential and the detuning in the adiabatic limit. The white lines mark the Andreev bound state energies. The parameters for the plot are $\GS/U$=0.3, $\GN/U=5\times10^{-3}$, $\kBT/U=10^{-2}$ and $\varphi=0$}
	\label{fig::APdeltamu}
\end{figure}

In the regime of adiabatic driving, the changes in the system parameters are much slower than the dynamics of the quantum dot such that we can approximate the quantum dot state at a given time $t$ by the stationary solution of the master equation with parameters corresponding to that time $t$. Hence, the driving frequency must satisfy $\omega\ll\Gamma/\hbar$. In addition, the amplitude of the parameter variation has to be sufficiently small such that  $\omega\delta_1\ll (k_\text{B}T)^2$ or $\omega\GS\ll (k_\text{B}T)^2$ hold, respectively~\cite{splettstoesser_adiabatic_2008}. Since non-Markovian memory effects can be neglected in the adiabatic regime, the system is effectively in an instantaneous stationary state at any time which implies that the dynamic properties in the adiabatic regime can be deduced from the static properties of our setup.

In Fig.~\ref{fig::APdeltamu}, we have plotted the dot's pair amplitude as a function of the detuning $\delta$ and the chemical potential of the normal lead $\mu_\text{N}$ for a fixed phase difference $\varphi=0$. We find that there are parameter regimes where the pair amplitude of the dot vanishes exactly. For negative detuning, this happens for chemical potentials $\mu_\text{N}$ between the Andreev bound state energies $E_{\text{A},+,-}$ and $E_{\text{A},-,-}$ while for positive detuning, this occurs for $E_{\text{A},+,+}>\mu_\text{N}>E_{\text{A},-,+}$. For these parameters, the dot is preferably singly occupied such that no superconducting proximity effect can take place. For other parameters, the pair amplitude is in general finite. It takes particularly large values of up to $|\mathcal F|\approx 0.25$ close to the particle-hole symmetric point $\delta=0$ where the proximity effect is strongest.

By inspecting Fig.~\ref{fig::APdeltamu}, we see that there are different ways of driving pronounced oscillations of the pair amplitude in the adiabatic regime. One can, e.g., fix the chemical potential of the normal lead at $\mu_\text{N}\approx \pm U/2$ and drive the pair amplitude smoothly via the detuning. Alternatively, one can fix the detuning around $\delta=\pm U/2$ and drive the pair amplitude via a modulation of the electrochemical potential $\mu_\text{N}$. As the electrochemical potential crosses the Andreev bound states, this leads to abrupt changes in the dot's pair amplitude.

\begin{figure}
	\includegraphics[width=\columnwidth]{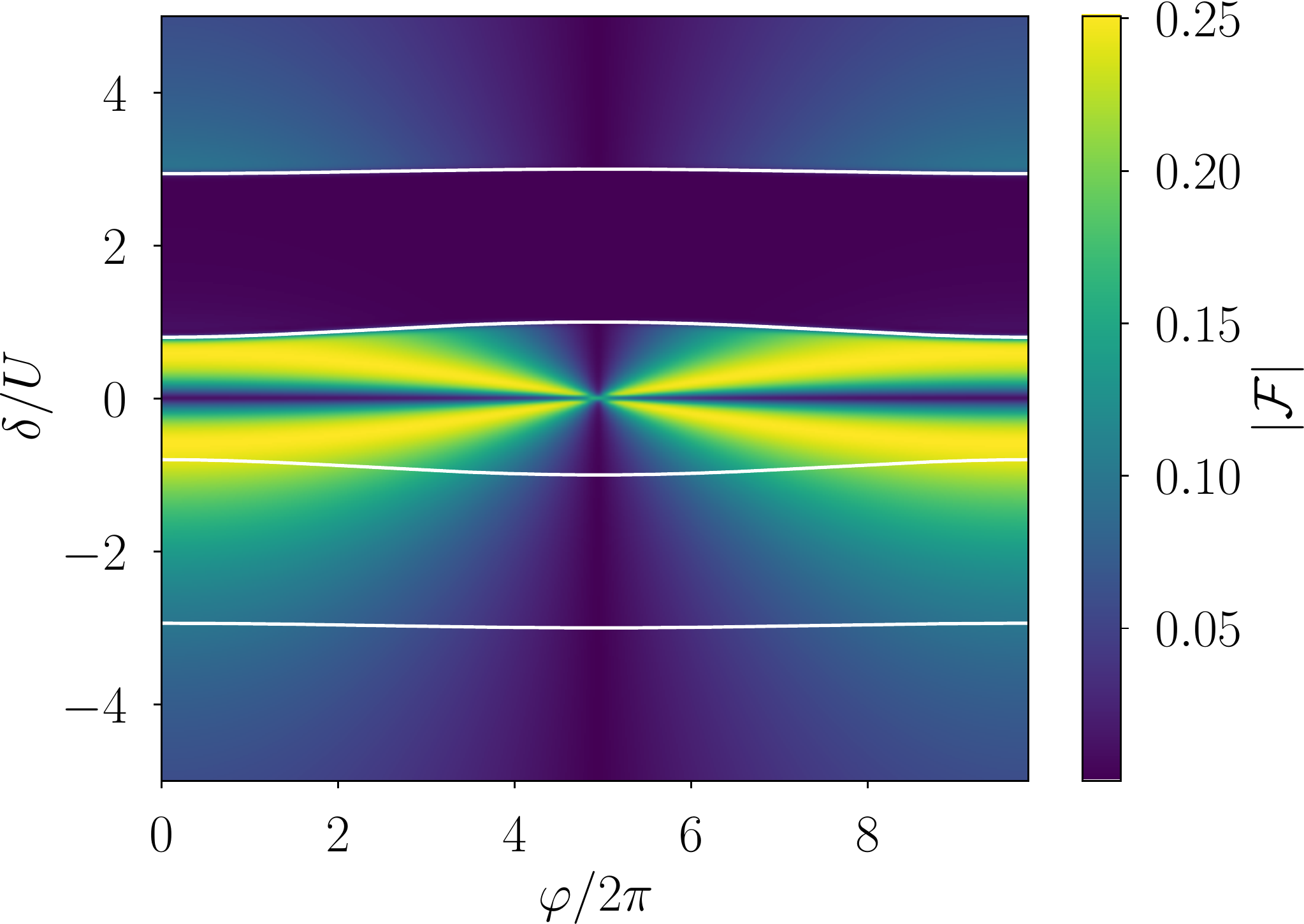}
	\caption{Pair amplitude as a function of the detuning $\delta$ and the phase difference $\varphi$ in the adiabatic limit. The white contours mark the Andreev bound state energies. The plot parameters are $\Gamma_S/U$=0.3, $\Gamma_N/U=5\times10^{-3}$, $\mu_N/U=1$, $k_BT/U=10^{-2}$ and $\varphi=\pi/4$}
	\label{fig::APdeltaphi}
\end{figure} 
Figure~\ref{fig::APdeltaphi} shows the pair amplitude as a function of detuning $\delta$ and phase difference $\varphi$ at a fixed electrochemical potential of the normal lead $\mu_\text{N}=U$. As discussed above, the pair amplitude $|\mathcal F|$ vanishes for detunings $E_{\text{A},-,+}<\delta<E_{\text{A},+,+}$ because the dot is singly occupied for these parameters. Similarly, the pair amplitude vanishes for a phase difference $\varphi=\pi$ as in this case, the accumulation term $\vec A$ and the exchange field $\vec B$ point along the $z$ direction such that no $x$ and $y$ component of the pseudospin are generated in the stationary state. For other phase differences, the dot exhibits again a sizeable pair amplitude if the detuning is chosen sufficiently small which opens up a way to drive the pair amplitude dynamics also via the superconducting phase difference.

Just as in the case of a quench, the pseudospin dynamics in the adiabatically-driven system can be accessed experimentally via the time-resolved charge currents in the system. Interestingly, we find that in the adiabatic regime, the Josephson current is much larger Andreev current. This results from the fact that the pseudospin accumulation generates only a small time-dependent $y$ component of the pseudospin which is about three orders of magnitude smaller than the pseudospin components in the $x-z$ plane. According to Eqs.~\eqref{eq::Ijos} and~\eqref{eq::Iand}, this immediately implies that the time-dependent contribution to the Andreev current is orders of magnitude smaller than the time-dependent Josephson current such that it is possible to reconstruct the pseudospin dynamics from the Josephson current alone.

\subsubsection{\label{sec:fast}Fast Driving}
\begin{figure}
	\includegraphics[width=\columnwidth]{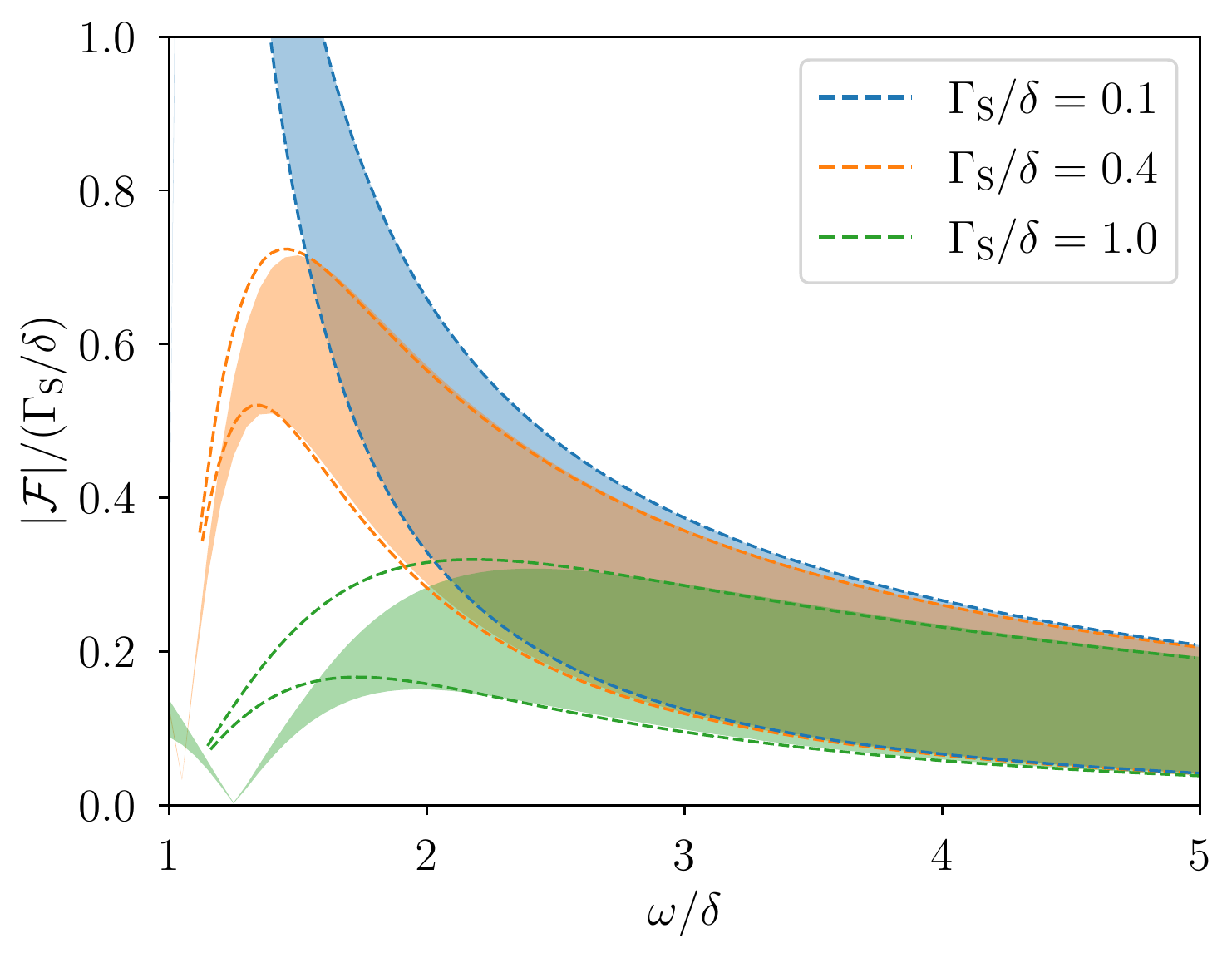}
	\caption{Range of the pair amplitude oscillation as a function of the external driving frequency in the fast-driving limit for different coupling strengths to the superconductors $\GS=\delta/10$ (blue), $\GS=0.4\,\delta$ (orange) and $\GS=\delta$ (green). The dashed lines mark the solution of our analytic approximation and the filled areas the solution of the numeric calculation. For each driving frequency, the plot marks the upper and lower bound of the pair amplitude. Plot parameters are $\GN=5\times 10^{-3}U$, $\delta=U$, $\mu_\text{N}=-10U$, $\kBT=U/100$.}
	\label{fig::FastPair}
\end{figure}

In the following, we discuss the opopsite limit when the system is driven with a frequency greater than the coupling strength to the superconductor. This limit is in particular relevant for driving the system via the phase difference $\varphi$. One can achieve a time-dependent phase difference via the AC Josephson effect by applying a voltage $V$ between the superconducting reservoirs which will give rise to a phase that changes on a time scale given by $\hbar/(eV)$ which can be larger than $\hbar/\GS$.
We tackle the problem in the limit of a large bias voltage applied to the normal conductor. On the one hand, this minimizes the relaxation rate in the system such that we can expect a large pair amplitude in this limit. On the other hand, if the chemical potential is much larger than the Andreev bound state energies, we can neglect level renormalizations which simplifies our analysis. Finally, transport becomes unidirectional in the large bias limit which further simplifies the problem.

In order to describe the fast driving regime, we make a Fourier ansatz for the pseudospin, $\mathbf{I}=\sum_n \mathbf{I}_n \exp(\iu\omega t)$, and split the exchange field into its Fourier components $\mathbf{B}=\mathbf{B}_0+\mathbf{B}_+ \exp(\iu \omega t)+\mathbf{B}_-\exp(-\iu \omega t)$ with $\mathbf{B}_+=\mathbf{B}_-=(\GS,0,0)^T$, $\mathbf{B}_0=(0,0,-\delta)^T$.
By inserting the Fourier ansatz into the master equation, we obtain an infinite hierarchy of equations for the Fourier components of $\vec I$,
\begin{equation}
	\mathbf{G}_n\mathbf{I}_n+\mathbf{B}_+(\mathbf{I}_{n-1}+\mathbf{I}_{n+1})=\delta_{0n}\mathbf{A} \label{eq::fast}
\end{equation} 
with $\mathbf{G}_n=(\GN+\iu n \omega)\mathbf{1}+\mathbf{B}_0$ and $\mathbf{A}=(0,0,\GN/2)^T$.
Equations of this type occur generically in the Floquet analysis of driven quantum systems and can be solved by the method of matrix-continued fractions~\cite{hanggi_driven_1998}. To this end, we introduce ladder operators for lowering $\mathbf{L}_n\mathbf{I}_n=\mathbf{I}_{n-1}$ and raising $\mathbf{R}_n\mathbf{I}_n=\mathbf{I}_{n+1}$ the Fourier index.
The ladder operators can be calculated from Eq.~\eqref{eq::fast} for the case of $n\neq0$. The result of the ladder operators are the continued fractions
\begin{align}
	\mathbf{R}_{n-1} &=-[\mathbf{G}_n+\mathbf{B}_+\mathbf{R}_n]^{-1}\mathbf{B}_+,\\
	\mathbf{L}_{n+1} &=-[\mathbf{G}_n+\mathbf{B}_+\mathbf{F}_n]^{-1}\mathbf{B}_+.
\end{align} 
The ladder operators allow us to solve Eq.~\eqref{eq::fast} for $n=0$. In the fast-driving limit, it is sufficient to take into account only the ladder operators $\mathbf{L}_{0}$ and $\mathbf{R}_{0}$. Formally, the continued-fraction method is a systematic expansion in $\GS/(\hbar \omega)$. When we take into account $\vec L_0$ and $\vec R_0$ only, we neglect the generation of higher harmonics. As has been discussed in Ref.~\cite{kamp_higgs-like_2021} for the pair amplitude dynamics of a quantum dot weakly coupled to superconductors with a finite gap, the amplitude of higher harmonics is suppressed as the quantum dot cannot follow the external driving at extremely high frequencies. Therefore, the pseudospin dynamics is well characterized by the Fourier components $\vec I_0$ and $\vec I_{\pm1}$.

In Fig.~\ref{fig::FastPair}, we show the range of the pair amplitude oscillation as a function of the driving frequency $\omega$. As can be seen in Fig.~\ref{fig::FastPair}, the agreement between our continued-fraction approximation (dashed lines) and the numerical solution of the full master equation (filled area) is good for sufficiently large driving frequencies. We find that the agreement between the approximative solution and numerics starts at lower driving frequencies the smaller the coupling to the superconductor is. Interestingly, the agreement between analytics and numerics does not improve if we take into account additional ladder operators and higher Fourier components.
As can be seen in Fig.~\ref{fig::FastPair}, the system is able to follow the external driving up to high frequencies that are larger than the coupling to the superconductors. For fast driving, the average pair amplitude scales with $(\GS/\omega)^2$. This is in strong contrast to the case of a weakly-coupled superconductor-quantum dot hybrid~\cite{kamp_higgs-like_2021} where the pair amplitude dynamics is strongly suppressed in the fast driving case. 

\section{\label{sec:conclusion}Conclusions}
We have theoretically analyzed the dynamics of the superconducting pair amplitude of a quantum dot strongly coupled to two infinite-gap superconductors and weakly coupled to a normal metal by means of a real-time diagrammatic approach that enables us to account for strong Coulomb interactions and the superconducting proximity effect in a nonequilibrium scenario. 

We found that the intrinsic dynamics of the pair amplitude after a quench is governed by two effects. On the one hand, there is a coherent oscillation of the pair amplitude on a time scale $\hbar/\GS$ due to the tunneling of Cooper pairs between the dot and the superconductors. On the other hand, the pair amplitudes decays exponentially on a time scale given by $\hbar/\GN$ due to the tunneling of single electrons between the dot and the normal metal. 

The damping can be overcome by driving the system externally. We analyzed the driven dynamics in the limit of adiabatic driving where the dynamics can be linked to the static properties of the system. Furthermore, we investigated the system's dynamics under fast driving by means of a Fourier analysis that revealed that the dynamics of the pair amplitude can follow the external drive up to frequencies significantly larger than the coupling strength to the superconductors.

Our results demonstrate that a strongly-coupled superconductor-quantum dot hybrid exhibits a rich pair amplitude dynamics. The tunability of quantum-dot systems allows for a straightforward manipulation of the dynamics. The link between charge currents and pair amplitude opens up the possibility to detect the pair-amplitude dynamics experimentally. Our results pave the way for future studies of nontrivial pair-amplitude dynamics in other superconducting hybrid structures.

\acknowledgments
We acknowledge financial support from the Ministry of Innovation NRW via the \textquotedblleft Programm zur Förderung der Rückkehr des hochqualifizierten Forschungsnachwuchses aus
dem Ausland\textquotedblright\space and Deutsche Forschungsgemeinschaft (DFG, German Research Foundation) Project No. 278162697, SFB 1242.

\appendix
\section{\label{app::eigensystem}Eigensystem of the Master equation}
To solve the master equation from Eq.~\eqref{eq::bloch} we rewrite it in the form 
\begin{equation}
	\hbar\frac{d\mathbf{I}}{dt}=\mathbf{A}+\mathbf{M}\cdot\mathbf{I},\nonumber
\end{equation}
with matrix $\mathbf{M}$ representing the dynamics due to relaxation and the exchange field. For a symmetric coupling between the quantum dot and the superconductors the matrix is given by
\begin{equation}
	\mathbf{M}=\left(\begin{array}{ccc}
		-R_{xx} & B_{z} & -R_{xz} \\
		-B_{z} & -R_{xx} & B_{x} \\
		-R_{xz} & -B_{x} & -R_{zz} 
	\end{array}\right).\nonumber
\end{equation}
The solution of the master equation can be expressed in terms of the left and right eigensystem of $\vec M$. The eigenvalues $\lambda_i$ are given by the zeros of the characteristic polynomial 
\begin{align}
	0=&-2 B_x B_z R_{xz} - B_x^2(R_{xx}+\lambda)-B_z^2(R_{zz}+\lambda)\nonumber\\ &-(R_{xx}+\lambda)(-R_{xz}^2+(R_{xx}+\lambda)(R_{zz}+\lambda)).\nonumber
\end{align}
The associated unnormalized left and right eigenvectors are given by 
\begin{align}
	\mathbf{l}_j=\left(\begin{array}{c}
		B_z(B_x+R_{yz})+(B_y-R_{xz})(R_{xx}+\lambda_j) \\
		B_z(B_y-R_{xz})-(B_x+R_{yz})(R_{xx}+\lambda_j) \\
		B_z^2+(R_{xx}+\lambda_j)^2
	\end{array}\right),\nonumber
	\\
	\mathbf{r}_j=\left(\begin{array}{c}
		B_z(B_x-R_{yz})-(B_y+R_{xz})(R_{xx}+\lambda_j) \\
		B_z(B_y+R_{xz})+(B_x-R_{yz})(R_{xx}+\lambda_j) \\
		B_z^2+(R_{xx}+\lambda_j)^2
	\end{array}\right),\nonumber
\end{align}  
respectively.

% \bibliography{/home/bjoern/LaTeX/Bibtex/Meine_Bibliothek.bib}

%apsrev4-2.bst 2019-01-14 (MD) hand-edited version of apsrev4-1.bst
%Control: key (0)
%Control: author (8) initials jnrlst
%Control: editor formatted (1) identically to author
%Control: production of article title (0) allowed
%Control: page (0) single
%Control: year (1) truncated
%Control: production of eprint (0) enabled
%

\end{document}